\documentclass[twocolumn,twoside,showkeys,preprintnumbers,amsmath,amssymb,secnumarabic, nofootinbib, nobibnotes,epl]{revtex4}

\usepackage{epsfig}
\usepackage{graphicx}
\usepackage{fancyhdr}

\begin{document}


\title{Symbolic Sequences and Tsallis Entropy}

\author{H. V. Ribeiro}\email{hvr@dfi.uem.br}
\author{E. K. Lenzi}
\author{R. S. Mendes}
\affiliation{Departamento de F\'isica, Universidade Estadual de Maring\'a, \\ Av. Colombo 5790, 87020-900, Maring\'a, PR, Brazil}

\author{G. A. Mendes}
\author{L. R. da Silva}
\affiliation{Departamento de F\'isica, Universidade Federal do Rio Grande do Norte \\ 59072-970 Natal, RN, Brazil}


\begin{abstract}
We address this work to investigate symbolic sequences with
long-range correlations by using computational simulation. We
analyze sequences with two, three and four symbols that could be
repeated $l$ times, with the probability distribution $p(l)\propto
1/ l^{\mu}$. For these sequences, we verified that the usual
entropy increases more slowly when the symbols are correlated and
the Tsallis entropy exhibits, for a suitable choice of $q$, a
linear behavior. We also study the chain as a random walk-like
process and observe a nonusual diffusive behavior depending on the
values of the parameter $\mu$.
\end{abstract}

\maketitle
\thispagestyle{fancy}
\setcounter{page}{1}

\section{Introduction}

Two basic assumptions of the statistical mechanics are: the
``equal a priori probabilities'' and ergodicity. When these
assumptions do not hold, we need other suitable tools to study
systems which exhibit a nonusual behavior. A typical situation can
be found by analyzing systems which have an intermediate regime
between periodic  and chaotic \cite{Gaspard}. This kind of system
commonly shows a power law spectra and appears in several fields
of science. Aspects of nonusual behavior have been explored, for
instance, in biology\cite{Mendes}, nuclear physics \cite{Drago},
financial market \cite{Queiros}, music \cite{Borges2} and
linguistics \cite{Montemurro}. In this context, there are also
works that search for correlations in DNA sequences
\cite{Krishnamachari,Peng,Herzel,Li,Allegrini} by using entropic
indexes \cite{Ebeling,Li2,Herzel2,Schmitt,Lio}. To provide a
possible description for these systems which are not conveniently
explained by the usual formalism, Tsallis\cite{Tsallis} proposes
an extension of the Boltzmann-Gibbs entropy. Many systems have
been investigated by using this approach, e.g., long-range
Hamiltonian systems like the HMF model \cite{Pluchino}, the
generalized Lennard-Jonnes gas \cite{Borges}, self-gravitating
systems \cite{Du} and anomalous diffusion \cite{Assis}.

In this direction, to try to clarify in a more direct way basic
aspects related to Tsallis entropy, it may be convenient to
consider specifc models with a kind of long-range behavior. 
Considering this, the aim of this work is to explore the nonusual behavior
of a symbolic model with an adjustable long-range behavior. More
precisely, we investigate one dimensional symbolic sequences with
long-range correlations which are generated by using the numerical
experiment presented in Ref. \cite{Buiatti}. The procedure uses
two random numbers to obtain a lattice with $N$ sites which
represent the symbolic sequence. One of them, $x$, has a uniform
distribution in the interval [0,1] and the other emerges from the
expression
\begin{equation}
   y=A\left[ \frac{1}{(1-x)^{1/(\mu-1)}}-1\right]\,,
\end{equation}
where $A$ and $\mu$ are real parameters. We go through the
symbolic sequence drawing $x$ and filling $N_y=[y]+1$ sites with
the same value $z$, where $[y]$ denotes the integer part of $y$
and $z$ is a signal generator that can have one of four distinct
values (0, 1, 2, 3) with the same statistical weight. A typical
example obtained within this procedure is
\begin{equation}
 Q = \left\lbrace \underbrace{0,0,0,}_{N_y=3}\underbrace{1,1,1,1,1,}_{N_y=5}\underbrace{0,0,}_{N_y=2}
\underbrace{1,1,1,1,}_{N_y=4}\underbrace{0}_{N_y=1} \right\rbrace\,
\nonumber
\end{equation}
for a sequence with two symbols.

For the sequences generated with the procedure described above, we
may obtain the probability distribution function of the variable
$y$, $p(y)$, and show that, depending on the values of the $\mu$, it can be
asymptotically related to a L\'evy distribution (for $y\geq0$). In
fact, after some calculations, one can show that $p(y)$ is given by
\begin{equation}
\label{p1}
   p(y)=(\mu-1) \frac{A^{\mu-1}}{(A+y)^\mu}\;,
\end{equation}
and the first moment of this distribution is \mbox{$\langle
y\rangle=A/(\mu-2)$}. By comparing the asymptotic limit of
Eq.(\ref{p1}), $p(y)\sim 1/y^{\mu}$, with the asymptotic limit of
the L\'evy distributions, $p(y)\sim 1/y^{1+\eta}$, the relation
between $\mu$ and $\eta$ is $\mu=1+\eta$. Note also that $\langle
y\rangle$ diverges for $\mu\rightarrow2$. This fact indicates
that, when $\mu$ is close to two, $N_y$ may assume large values and
fill a large part of the symbolic sequence with the same symbol.
On the other hand, when $\mu$ is far from two ($\mu\gg2$), large
values of $N_y$ become very rare and consequently the sequence has
more alternated symbols.

\begin{figure*}[ht!]
   \begin{center}
      \includegraphics[scale=0.23]{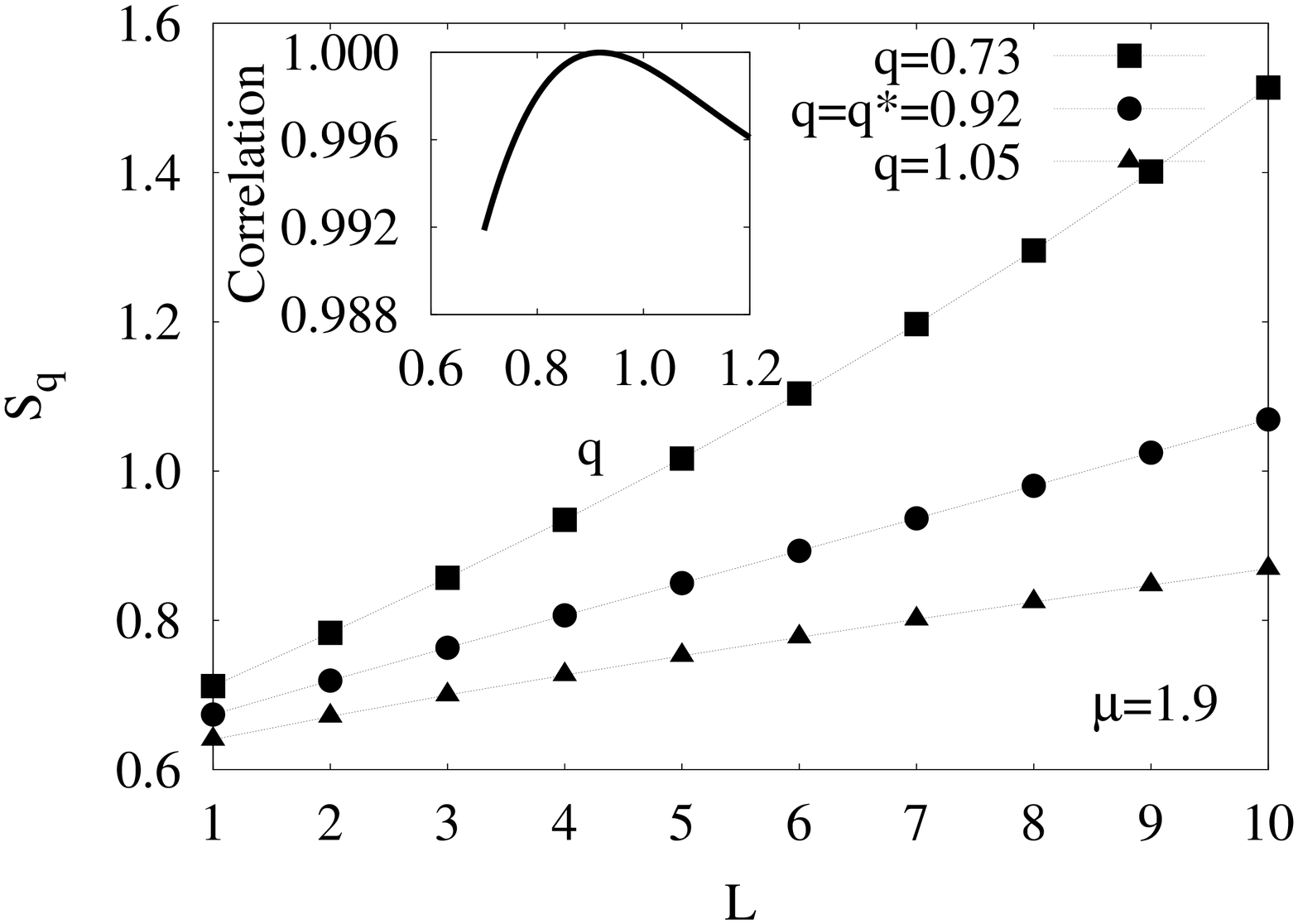}
      \includegraphics[scale=0.23]{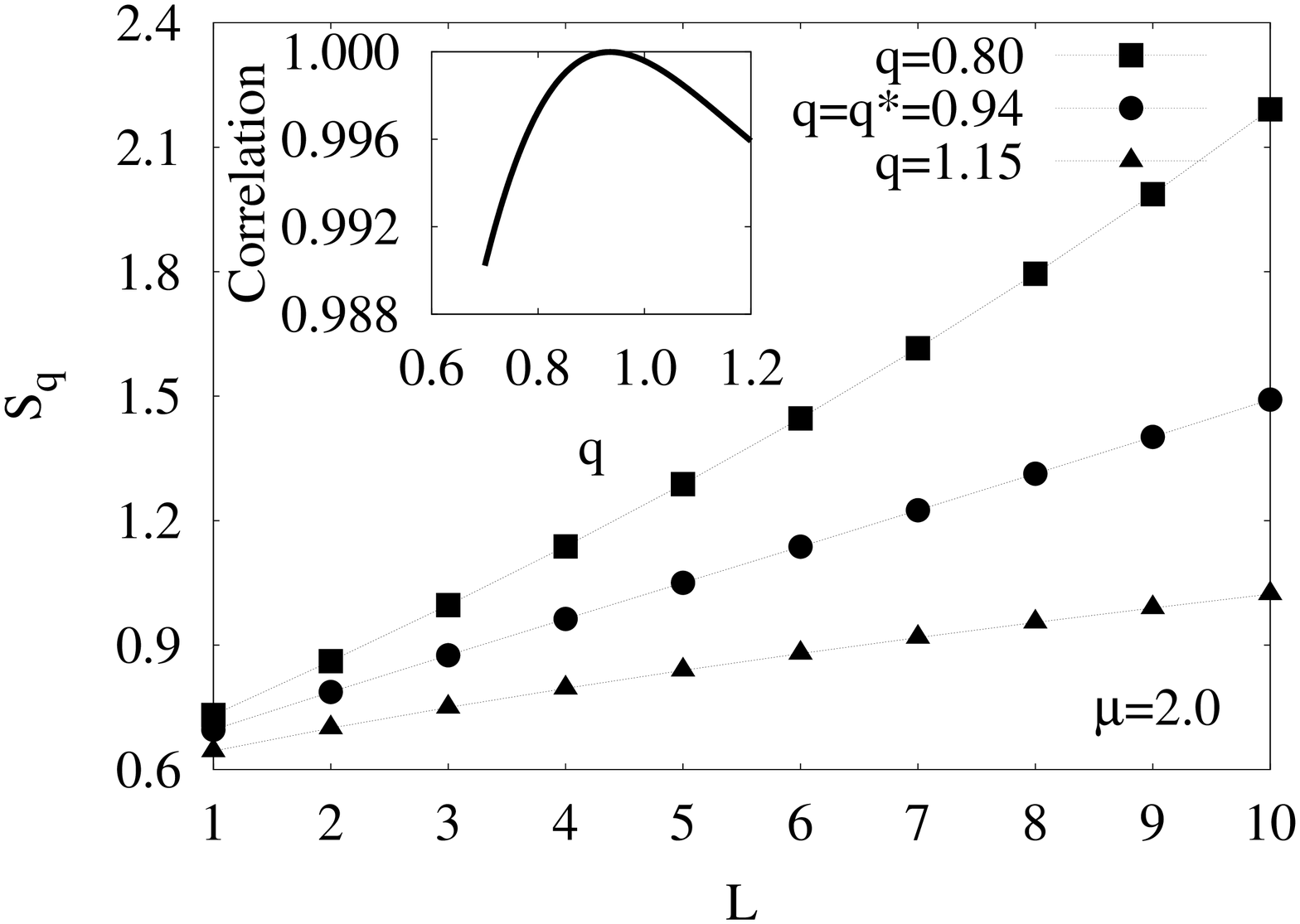}
      \includegraphics[scale=0.23]{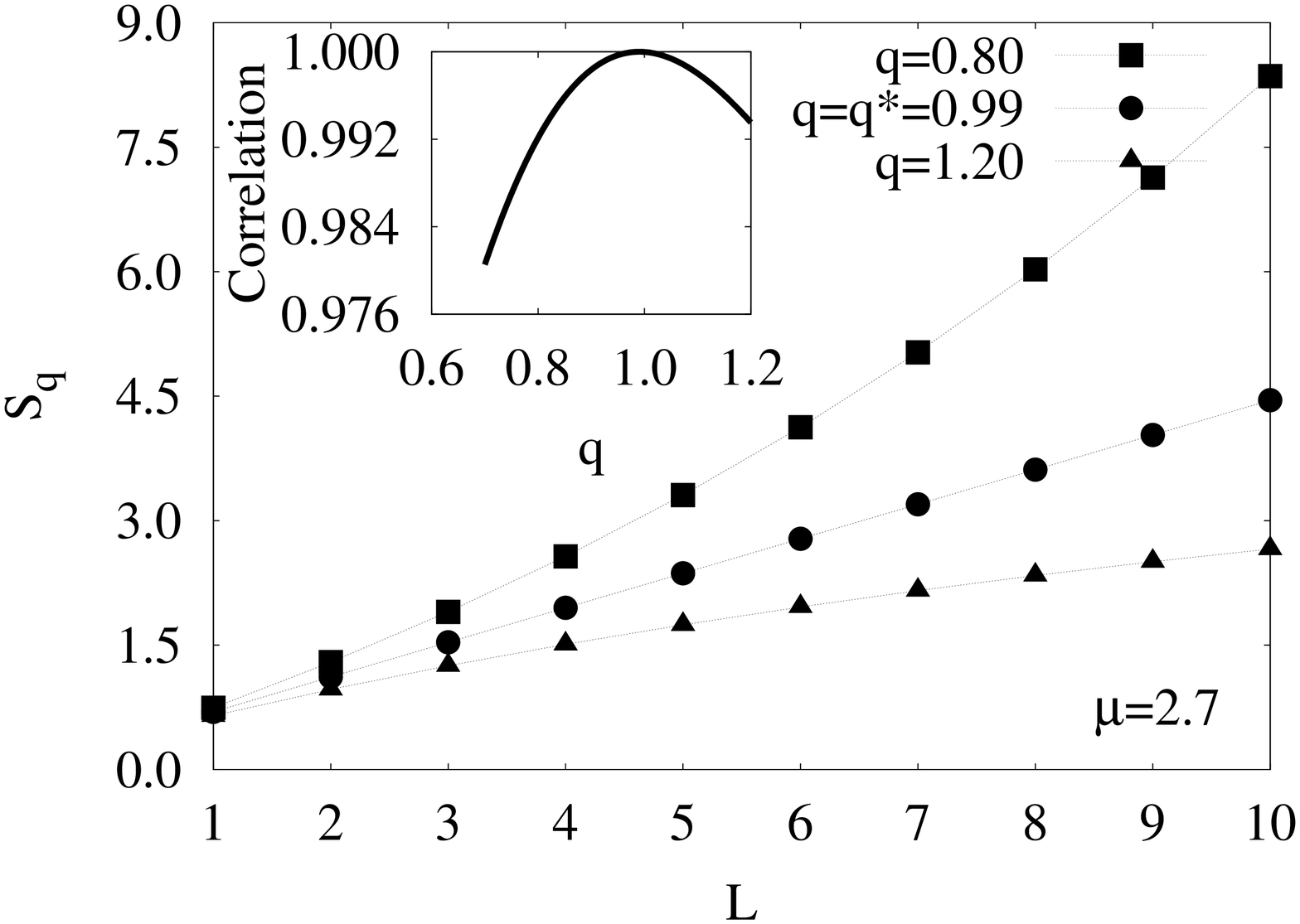}
      \caption{$S_q$ versus $L$ for some values of $\mu$ (indicated in the
      figure) for a two-symbol sequence. We use $A=2$ and $N=10^8$ in all the
      three figures.} \label{fig:S_versus_L}
   \end{center}
\end{figure*}

\section{Entropy and Sequence}

The Tsallis entropy is defined, for a system with $W$  microstates
and occupation probabilities $p_i$, as follows:
\begin{equation}
   S_q=\frac{1-\sum_{i=1}^{W} p_i^q}{q-1},
\end{equation}
where $q$ is a real parameter. In the limit $q\rightarrow1$ we
have the standard Boltzmann-Gibbs entropy. $S_q$ is extensive for
a composite system consisting of independent subsystems for $q=1$
and nonextensive for $q\neq1$; for this reason, $S_q$ is sometimes
referred to as nonextensive entropy. However, when we have
long-range interactions or long-range correlations, the subsystems
cannot be independent. In this case we will see that $S_q$ can be
extensive for a particular value of $q\neq1$.

In order to evaluate the Tsallis entropy, for the symbolic
sequence generated with the previous procedure, we fix windows of
length $L$ which are moved along the sequence. Then, we count how
many times a given configuration (string) occurs, determining the
probability $p_i$ of a specific configuration $i$. To illustrate
this procedure, suppose that we have the following sequence:
\begin{equation}
 Q = \left\lbrace 0, 0, 1, 0, 0, 1, 1, 1, 0, 0, 0, 1, 0, 0, 0, 0, 1, 0, 1, 1\right\rbrace\,,  \nonumber
\end{equation}
then we fix a window of length $2$ and move it along the sequence,
i.e., we have
\begin{equation}
 Q = \left\lbrace \underbrace{0, 0,}_1 \underbrace{1, 0,}_2 \underbrace{0, 1,}_3 \underbrace{1, 1,}_4 \underbrace{0, 0,}_5 \underbrace{0, 1,}_6 \underbrace{0, 0,}_7 \underbrace{0, 0,}_8 \underbrace{1, 0,}_9 \underbrace{1, 1}_{10} \right\rbrace\,,  \nonumber
\end{equation}
where the index below the keys indicates time steps of the window's motion. The
next step is to count how many times a given configuration occurs,
for example, the configuration $\{0,0\}$ occurred $4$ times (in
the instants of ``time'' $1$, $5$, $7$ and $8$), leading to the
probability $4/10$. Similarly, we calculate the probability of
other configurations and for other window lengths as well.

Figure (\ref{fig:S_versus_L}) shows $S_q$ as a function of $L$ for
some values of $\mu$. Note that for each value of $\mu$ there is
only one value of $q=q*$ that makes the relation $S_q$ versus $L$
linear. This feature becomes evident when we look at the linear
correlations (see the insets in Fig. (\ref{fig:S_versus_L})). We
can observe from the above results that when $\mu$ decreases $q*$
also decreases.

\begin{figure}[ht!]
   \begin{center}
      \includegraphics[scale=0.58]{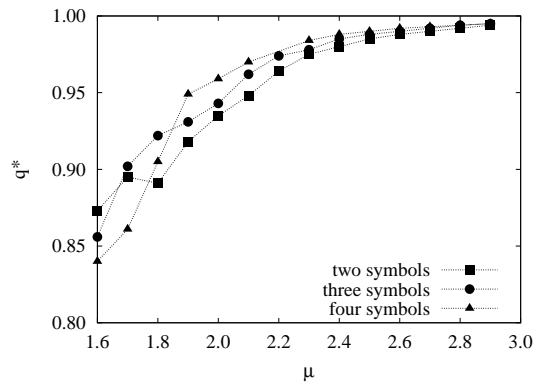}
      \caption{The entropic index $q*$ versus $\mu$ for two, three and
      four-symbol sequences, with $A=2$ and $N=10^8$.}
      \label{fig:q*_versus_mu}
    \end{center}
\end{figure}

Motivated by the previous results, we investigate the relation
$q*$ versus $\mu$ for two, three and four-symbol sequences. The
results are shown in Fig. (\ref{fig:q*_versus_mu}). Note that,
when $\mu$ increases, $q^{*}$ tends to unity, and that the more
symbols the sequence has, the faster it reaches towards one. This feature
shows that large values of $\mu$ generate small values of $N_y$
and consequently the terms of the symbolic sequence becomes
noncorrelated leading to the usual description based on the
Boltzmann-Gibbs entropy. However, when $\mu$ decreases, $N_y$  is
generally very large (remember that, when $\mu<2$, all the moments
of $p(y)$ diverge) and introduces correlation among the terms of
the symbolic sequence which are not properly described by the
usual formalism. The decreasing values of $q*$ reflects this
nonusual behavior. We emphasize that in this case the
Tsallis entropy is extensive and Boltzmann-Gibbs entropy is not,
indicating the applicability and robustness of the generalized
entropy.

\section{Diffusion and Sequence}

In order to explore further aspects of a symbolic sequence, let us
consider it as an erratic trajectory and establish a
correspondence with a diffusive process. For the case of
two-symbol sequences, we associate the symbol ``0'' with a jump of
unit length to the right and the symbol ``1'' with a jump of unit
length to the left. That is, a random walk-like process.

\begin{figure}[!t]
    \begin{center}
       \includegraphics[scale=0.58]{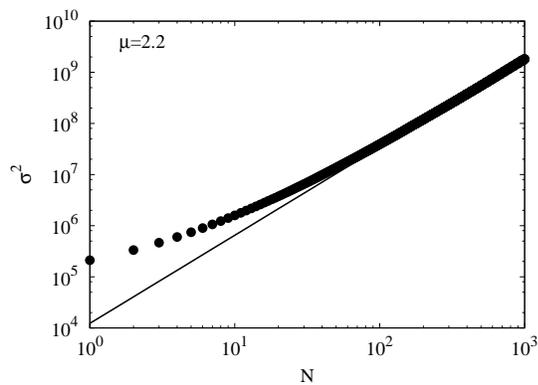}
       \caption{The standard-deviation versus $N$ for $\mu=2.2$ and $A=0.1$.}
       \label{fig:sd}
    \end{center}
\end{figure}

\begin{figure}[!t]
   \begin{center}
      \includegraphics[scale=0.56]{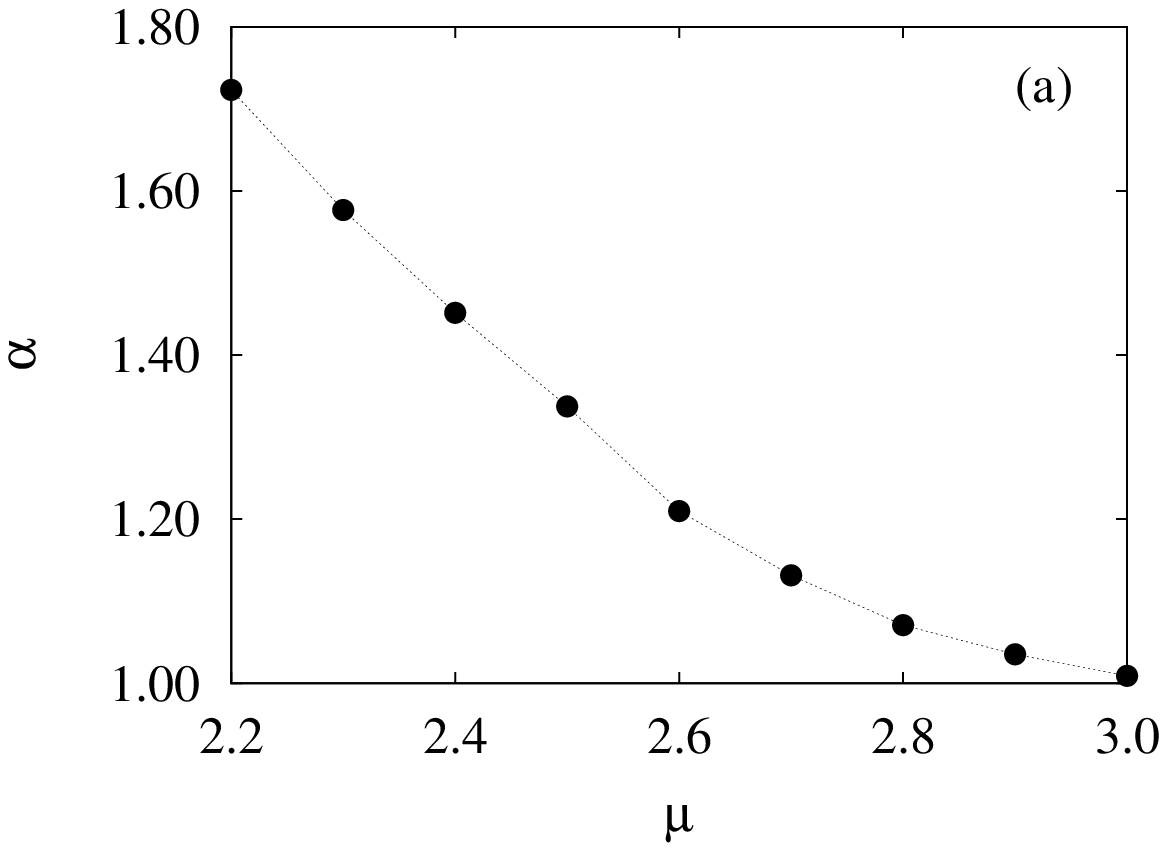}
      \includegraphics[scale=0.56]{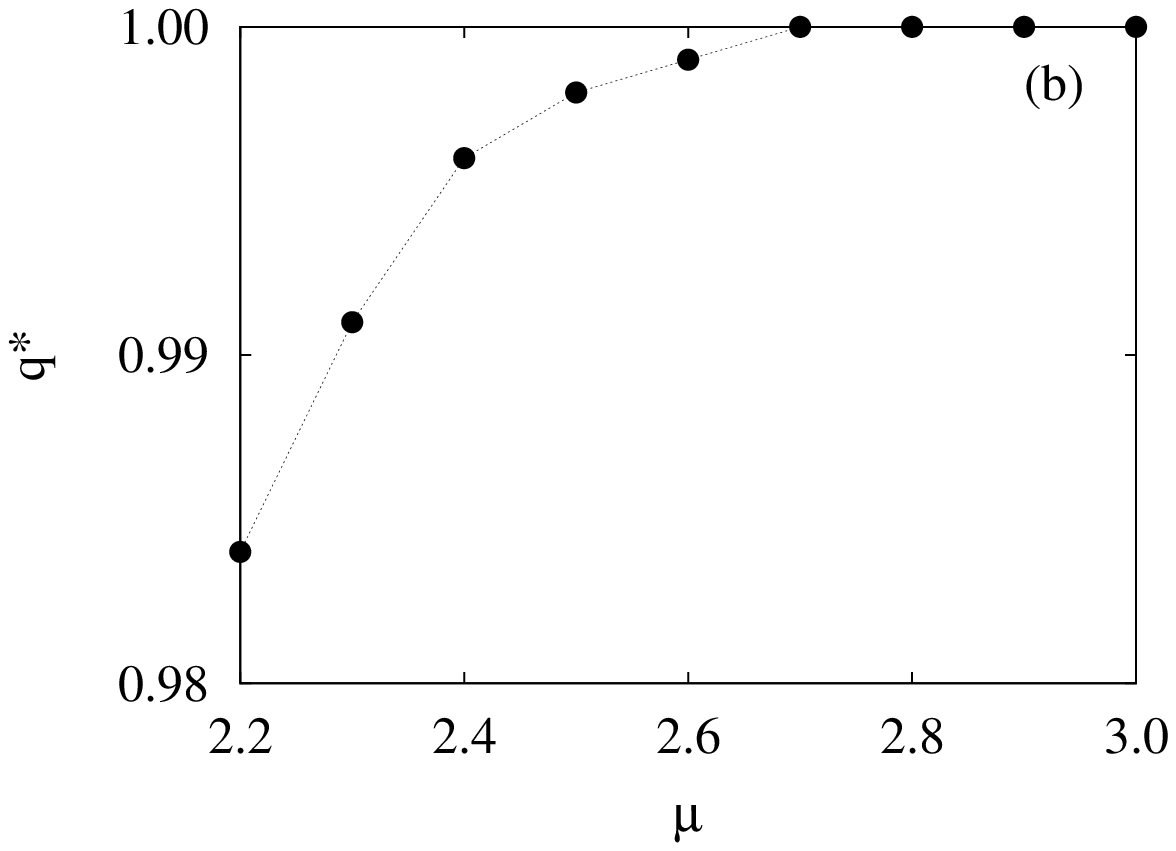}
      \includegraphics[scale=0.56]{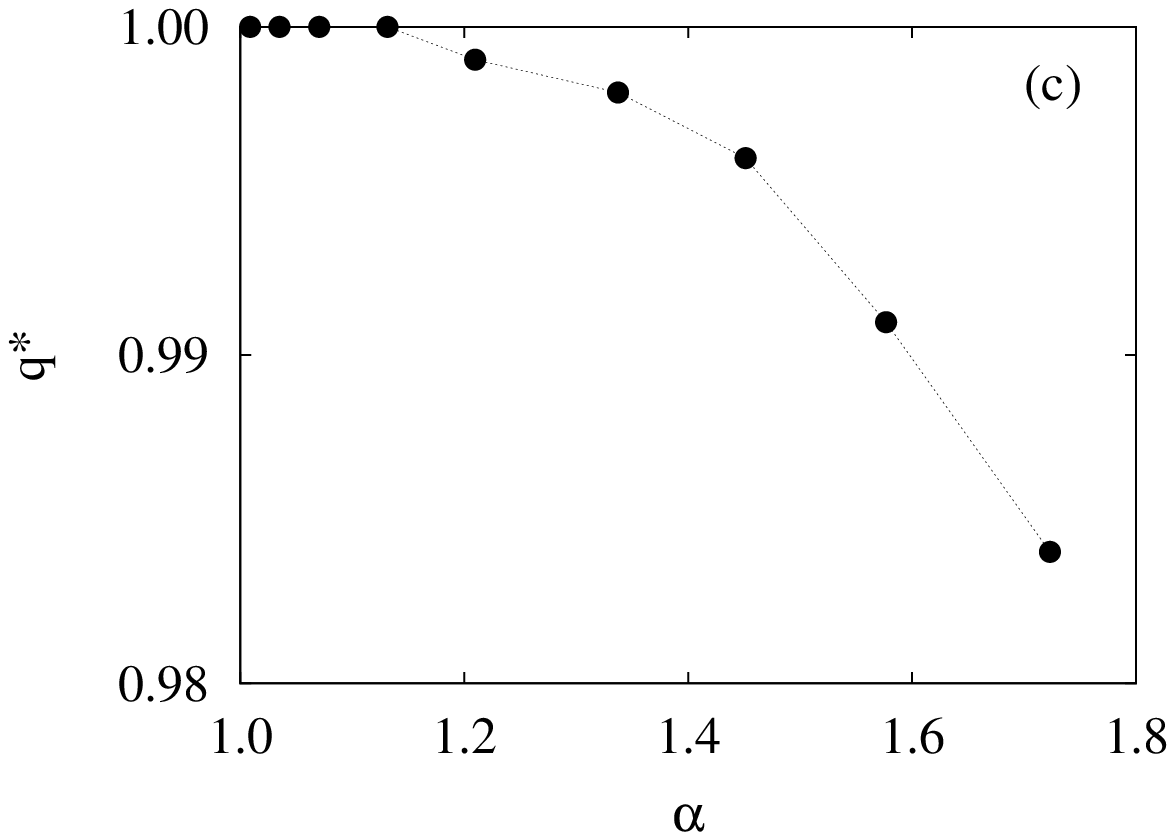}
      \caption{(a) The $\alpha$ slope of the curve in Fig (\ref{fig:sd}) versus $\mu$
               (b) the entropic index $q*$ versus $\mu$ and (c) $q*$ versus $\alpha$.
               In the three figures we use $A=0.1$.}
      \label{fig:mu_versus_alpha}
   \end{center}
\end{figure}

Using the previous prescription, we calculate the
standard-deviation for $i=1$ to $N$ over $10^5$ events as we can
see in Fig. (\ref{fig:sd}). We know that the slope $\alpha$ of
this curve is one for a usual diffusion, but in the case of
Fig.(\ref{fig:sd}) $\alpha$ is greater than one. We also observed
that $\alpha$ depends on $\mu$. This behavior is shown in
Fig.(\ref{fig:mu_versus_alpha}a).
\vspace*{0.4cm}
Note that for small values of $\mu$ ($\mu<3$) the diffusion is
anomalous, i.e., we have a superdiffusion, and for large
values ($\mu\geq3$) the diffusion regimes tend to a usual
diffusion. This behavior can be explained if we remember that for
small values of $\mu$, $N_y$ can be very large and consequently
the walker can make large steps without changing the direction.
When $\mu$ is large, this event becomes very rare because $N_y$ is
in general small, making the walker change directions,
producing a usual erratic trajectory. We may also connect $\alpha$
with $q*$ through the values of $\mu$. In order to do this we
evaluate the relation $q*$ versus $\mu$ as shown in Fig.
(\ref{fig:mu_versus_alpha}b) and exhibit $q*$ versus $\alpha$ in
Fig. (\ref{fig:mu_versus_alpha}c).

\section{Discussion and Conclusion}

We verified that by varying the value of $\mu$ we can produce
long-range correlations in symbolic sequences. This is evidenced
by the nonlinear growing of the Boltzmann-Gibbs entropy. This feature
led us to use the Tsallis entropy with suitable values of $q$ to
obtain a satisfactory description of these sequences.
Specifically, we observed that the Tsallis entropy preserves the
extensivity even when the terms of the symbolic sequence are
correlated. We also considered the symbolic sequence as a random
walk-like process and evaluated the standard deviation. The result
showed that the diffusive process presents a superdiffusive regime
which emerges for small values of $\mu$ ($\mu<3$). The usual
diffusion is recovered when $\mu\geq 3$.

\bigskip
\noindent
{\bf Acknowledgements}
\medskip

The authors thank CENAPAD-SP  (Centro Nacional de Processamento de
Alto Desempenho em S\~ao Paulo) for the computational support and
the CNPq/INCT-SC for financial support.

\end{document}